\documentclass[12pt]{article}

\usepackage{graphicx} 

\begin{document}
\thispagestyle{empty}
\begin{center}

\null
\vskip-1truecm
\vskip2truecm
{\bf THE COVARIANT ENTROPY BOUND, BRANE COSMOLOGY, AND THE NULL ENERGY CONDITION\\}
\vskip2truecm
Brett McInnes
\vskip2truecm

Department of Mathematics, National University of Singapore, 10 Kent Ridge Crescent,
Singapore 119260, Republic of Singapore.\\ 
E-mail: matmcinn@nus.edu.sg\\    

\end{center}
\vskip1truecm
\centerline{ABSTRACT}
\baselineskip=15pt
\medskip
In discussions of Bousso's Covariant Entropy Bound, the Null Energy Condition is always assumed, as a sufficient {\em but not necessary} condition which helps to ensure that the entropy on any lightsheet shall necessarily be finite. The spectacular failure of the Strong Energy Condition in cosmology has, however, led many astrophysicists and cosmologists to consider models of dark energy which violate {\em all} of the energy conditions, and indeed the current data do not completely rule out such models. The NEC also has a questionable status in brane cosmology: it is probably necessary to violate the NEC in the bulk in order to obtain a ``self-tuning" theory of the cosmological constant. In order to investigate these proposals, we modify the Karch-Randall model by introducing NEC-violating matter into $AdS_5$ in such a way that the brane cosmological constant relaxes to zero. The entropy on lightsheets remains finite. However, we still find that the spacetime is fundamentally incompatible with the Covariant Entropy Bound machinery, in the sense that it fails the Bousso-Randall consistency condition. We argue that holography probably forbids all {\em cosmological} violations of the NEC, and that holography is in fact the fundamental physical principle underlying the cosmological version of the NEC. 
\vskip3.5truecm
\begin{center}

\end{center}

\newpage

\addtocounter{section}{1}
\section*{1. The Status of the Null Energy Condition}
One of the major themes of recent theory has been the concept of holography, which originated from considerations of black hole entropy. This aspect of holography has been developed into the {\em Covariant Entropy Bound}, introduced by Bousso in \cite{kn:bousso1}. [We shall mostly follow the recent review paper \cite{kn:bousso}]. The entropy bound is particularly important as a testing ground for holographic ideas because it has been articulated with great precision and clarity.

Previous discussions of the entropy bound have assumed the validity of the Null Energy Condition [NEC], which just requires that for any null vector $k^a$, the stress tensor should satisfy
\begin{equation} \label{eq:one}
T_{ab}k^ak^b \geq 0.
\end{equation}
In general, this condition is not particularly well-motivated, unless one is also assuming the Weak Energy Condition [which implies the NEC by continuity]; the physical meaning of ``observers" moving at the speed of light is dubious indeed. But it is rather reasonable here. For without it we must expect, {\em in general}, to find zero-vorticity null congruences with permanently negative expansion, and this would generically lead to ``lightsheets" \cite{kn:bousso} with infinite entropy. Until very recently, in any case, few doubted the universal validity of the NEC [except perhaps in very exotic circumstances, remote from observations \cite{kn:visserworm}]. Furthermore, it was generally thought that there are strong theoretical reasons for doubting the existence of NEC-violating matter.

The sensational failure of the Strong Energy Condition revealed by the discovery of cosmic acceleration \cite{kn:carroll} has, however, changed many attitudes to energy conditions. Recently, old arguments for the NEC have been challenged \cite{kn:vissertwilight}\cite{kn:mcinnessmash}\cite{kn:mcinnesastro}\cite{kn:axel2} and the observational situation has been reviewed more critically \cite{kn:caldwell}. As the full extent of the challenge to the NEC does not appear to have been widely appreciated, let us very briefly survey the current position. 

There is now convincing observational evidence that the density of the baryonic and dark matter in our universe is far smaller than the critical density. Yet the many successes of the CDM model provide, along with the cosmic microwave data obtained in recent years, an almost equally convincing argument that the universe is spatially flat. General Relativity then implies that our world must contain vast quantities of ``dark energy" with a positive density which reconciles the two sets of observations. The celebrated supernova observations merely confirm this conclusion. [See \cite{kn:ratra} for a cautious review of the observations.]

None of this is very strange; the fact that we cannot ``see" most of the contents of our universe is regrettable but not unexpected. What {\em is} strange is the time evolution of the density of the dark energy. At the time of the synthesis of the light elements, its density must have been negligible compared with the density of radiation. Yet today it dominates all other forms of matter and energy combined. The usual inference is that the dark energy has an extraordinary property: its density does not change as the universe expands. [Alternatively, it is assumed that its density declines very slowly, as in quintessence models; see \cite{kn:ratra}.] This is the cosmological constant model. But if we are willing to entertain such a strange possibility, we should also be willing to consider the possibility that the dark energy density actually {\em increases} as the universe expands. This is, after all, the most obvious interpretation of the facts: the dark energy density was negligible in the past, is dominant now, and presumably --- unless we are privileged to live at a very special point in time --- will be still more dominant in the future.

In fact, Frampton \cite{kn:frampton} has recently argued that exotic dispersion relations [motivated by ``stringy dark energy" cosmologies \cite{kn:mersini}, descendants of the well-known Brandenberger-Vafa cosmologies \cite{kn:vafa}], naturally lead to an increasing density for the dark energy. While the basis of the particular model discussed in \cite{kn:frampton} is somewhat controversial \cite{kn:uzan}, the basic idea that non-standard dispersion relations can lead naturally to an increasing density is worth exploring, even if it is true that a realistic model requires fine-tuning. To see the consequences of this, consider the equation expressing the vanishing of the divergence of the dark energy stress tensor \cite{kn:ellis}:
\begin{equation} \label{eq:two}
{{d\rho}\over{dt}} + 3(\rho + p){{da}\over{dt}}/a = 0.
\end{equation}
This may be written as
\begin{equation}\label{eq:three}
{{d\,ln(\rho)}\over{d\,ln(a)}} = -3(1 + w),
\end{equation}
where $w$ is the dark energy equation-of-state parameter, the ratio of its pressure to its density, and $a$ is the scale factor. Thus we see that the density is constant if $w = -1$, the case of a cosmological constant, and that it increases as the universe expands only if $w < -1$. Thus the idea that the dark energy density is increasing requires us to take such values seriously. 

While Frampton's theory appears to be the first string-based model actually to require $w < -1$ as a matter of basic principle, violations of the NEC have in fact been proposed in quite a number of recent works. We shall now give a very compressed and incomplete survey of this literature. The intention here is {\em not} to persuade the reader that the NEC is violated either in cosmology or otherwise, but merely to demonstrate that such violations are not completely out of the question. The relevant works can be roughly categorised as follows.

First, one of the earliest instances of a challenge to the NEC arose in connection with inflation, where, in some versions, upward fluctuations of the Hubble parameter are required. The Einstein equations imply that such fluctuations must violate the NEC. See \cite{kn:guth} for a discussion of this.

Second, scalar fields have of course long played an important role in cosmology, and non-minimally coupled scalars can easily lead to interesting violations of the NEC. See for example \cite{kn:boisseau}\cite{kn:visser}\cite{kn:faraoni}\cite{kn:torres}, and also \cite{kn:arias} for a different approach.

Third, it is well known that quantum effects can lead to local violations of the NEC; see \cite{kn:olum}\cite{kn:ford}\cite{kn:fewster} for a sample of recent important work, and note that some of this work calls into question the widespread belief that quantum violations of the NEC are necessarily tightly constrained, particularly in higher dimensions. Violations of the NEC arising in quantum field theory have been directly applied to cosmology in several papers by Parker and collaborators \cite{kn:parker}; this work is notable for its specific prediction that $w$ should indeed be less than $-1$. See also \cite{kn:onemli}. Quantum gravity theories can also lead to such phenomena: see for example \cite{kn:bojowald}, which requires strong violations of the NEC in the remote past [though not at present].

Fourth, [at least temporary] violations of the NEC are required in many models which attempt to replace the Big Bang with a [smooth] ``bounce". Much of this work has been stimulated by the ``cyclic" and ``ekpyrotic" theories \cite{kn:turok}\cite{kn:seiberg}. See \cite{kn:noh}\cite{kn:pintoneto} for a sample of this literature. It is particularly important to note that bounces can arise from {\em effective} violations of the NEC when a generalised action involving higher-derivative terms is used. This point is explored in a major work of Tsujikawa, Brandenberger, and Finelli \cite{kn:brandenberger}.

Fifth, NEC violations appear quite naturally in braneworld cosmologies: see \cite{kn:myung} and its references for examples. The $w < -1$ aspect is discussed explicitly in \cite{kn:sahni}. Recently Dabrowski and co-workers \cite{kn:dabrowski1}\cite{kn:dabrowski2} have made the claim that realistic braneworld cosmologies actually {\em demand} $w < -1$. A very important development in this line is the work of Kachru and McAllister \cite{kn:kachru}, who discuss an NEC-violating braneworld model in the context of the Klebanov-Strassler solution of a Calabi-Yau compactification of II B string theory. Whether or not it is realistic, this example is important because it is often argued against NEC-violating theories that the underlying matter fields violate unitarity. That is indeed the case for {\em very simple} matter models, such as a single scalar field, but the Kachru-McAllister example shows that it may not be so in more complicated cases. The point is that apparent misbehaviour of an effective four-dimensional theory can be alleviated at a higher level. Notice too that this example indicates that string theory is not completely incompatible with violations of the NEC.

Finally, and of course most importantly, the observational evidence for the Cosmological NEC [that is, the statement that $w \geq -1$] is not entirely convincing, as Caldwell \cite{kn:caldwell} noted in a seminal paper; see also \cite{kn:hannestad}\cite{kn:lewis}\cite{kn:dicus}\cite{kn:magueijo}\cite{kn:krauss} for observationally-oriented papers in which the possibility that $w$ could be less than $-1$ is considered. One very recent analysis \cite{kn:guzzo} of the data claims, {\em very tentatively}, that they support the NEC; another \cite{kn:melchiorri}, almost simultaneous, claims that $w < -1$ is still a very real possibility. In short, the NEC is {\em not}, as is widely believed, well supported by the observations: the data are, at present, inconclusive. 

We can summarise all this very simply: {\em there are good theoretical and observational reasons to question the universal validity of the NEC}. This does {\em not} mean that the NEC, or its cosmological version, should be abandoned forthwith. What it does mean is that the truth of the NEC is not a trivial matter; it is no longer acceptable merely to assume it without further comment. 

In this work, we shall be concerned exclusively with the {\em Cosmological} NEC. As is emphasised in \cite{kn:guzzo}, the distinction is important, because the existence of NEC-violating matter at the cosmological level does not imply that we can necessarily isolate or manipulate ``chunks" of such matter so as to construct exotic objects such as wormholes. Conversely, it may be possible to give theoretical arguments which support the {\em Cosmological} NEC, without implying that the NEC is universally valid. If it is eventually confirmed that the data point to $w \geq -1$, this will mean that the Cosmological NEC is an observational fact requiring a theoretical explanation, just as the cosmic acceleration is a fact requiring explanation. We propose to seek this explanation in holography, in the particular guise discussed by Bousso in \cite{kn:bousso}. 

While the NEC is, as explained earlier, very useful in discussions of the entropy bound, it does not of course follow that the NEC is absolutely {\em necessary} here. In some cases, the NEC-violating matter could be such that the entropy on lightsheets is still always finite. If such examples can be found, then perhaps [a suitably modified version of] the entropy bound need no longer rely on the NEC. We shall present an extremely simple example of this kind. It is motivated by the ``self-tuning" brane cosmologies. [See \cite{kn:carrollmersini} for a recent discussion and for references.] In these models, the cosmological constant on the braneworld is automatically ``tuned away" by the bulk, and the brane becomes flat. It has been shown by Cline and Firouzjahi \cite{kn:cline1}\cite{kn:cline2} that the most promising versions of this idea actually require a violation of the NEC in the bulk [granted that it is not violated on the brane], so this is quite a natural context for considering NEC-violation. [Another unusual, and possibly related, form of matter in the bulk is the Casimir energy. See \cite{kn:odintsov} and its references.] We stress, however, that we are {\em not} proposing this as even a quasi-realistic braneworld model. We shall not discuss the binding of the graviton to the brane, the effect of  more complicated models of NEC-violating matter, whether there are interesting perturbations away from exact flatness of the brane, and so on. Our objectives are more basic: we simply wish to determine whether this form of holography can make sense in the presence of very simple NEC-violating matter.

The example is obtained, after a brief explanation of the usual motivation for using the NEC, by simply introducing NEC-violating matter into $AdS_5$ in such a manner that the {\em negative} cosmological constant on the well-known Karch-Randall brane \cite{kn:karch1}\cite{kn:porrati} is relaxed to zero. A numerical investigation shows that the entropy on lightsheets emanating from the brane remains finite, confirming that the NEC is not necessary to ensure this. Nevertheless, in the following section we show that the holographic formulation of this situation is not just physically unrealistic --- it is internally inconsistent, by the Bousso-Randall consistency condition \cite{kn:boussorandall}. The meaning of this is discussed in the Conclusion.

\section*{2. The Raychaudhuri Equation and Lightsheets}

Given a surface of $n - 1$ spatial dimensions and area $A$ in an $n + 1$ dimensional spacetime, the n dimensional null hypersurfaces corresponding to null geodesics pointing directly towards directions in which the {\em expansion}, $\theta$, is non-positive, are called \cite{kn:bousso} {\em lightsheets}. The Covariant Entropy Bound states that the entropy on any lightsheet is bounded by $A/4$ in Planck units. Obviously it is desirable that lightsheets should terminate at a finite distance, since otherwise we are in danger of having infinite entropy on the lightsheet. [Actually this does not necessarily follow, since the geometry may be such that infinite distance corresponds to finite volume, as in the case of the Poincar\'e patch of AdS. Even if the volume {\em is} infinite, the relevant energy density may decay towards infinity --- in fact this must be the case if we are to obtain an asymptotically AdS spacetime --- and so infinite volume might correspond to finite energy and perhaps finite entropy. Usually, however, the volume elements of asymptotically AdS spaces increase too rapidly for this to save the day.]

By definition, the lightsheet terminates when the expansion is about to become positive. The null version of the Raychaudhuri equation [\cite{kn:wald}, page 222],
\begin{equation} \label{eq:four}
{{d\theta}\over{d\lambda}} = {{{-1}\over{n-1}}{\theta^2}} - \sigma_{ab}\sigma^{ab} + \omega_{ab}\omega^{ab} - 8\pi T_{ab}k^{a}k^{b},
\end{equation}
where $\lambda$ is an affine parameter for a null geodesic congruence, $k^a$ is the tangent vector, $\sigma_{ab}$ is the shear, and $\omega_{ab}$ is the vorticity, shows that lightsheets {\em must} terminate at a finite value of the affine parameter, provided that the vorticity is zero and the NEC is satisfied. [The equation governing the evolution of the vorticity [\cite{kn:wald}, page 223] ensures that if the congruence is initially hypersurface-normal, then the vorticity is permanently zero, so we can ignore this term.] For in that case, the right side is clearly negative if the initial value of $\theta$ is non-zero, and it is easy to show that $\theta$ tends to $-\infty$ at a finite value of $\lambda$. Again, for a generic geometry, this means that all lightsheets end on a caustic at a finite distance from the initial area.  If the NEC is {\em not} satisfied away from the initial surface, then there are various possibilities. 

Firstly, of course, the first two terms on the right hand side may still dominate, and a caustic may still form. The entropy on the lightsheet will then still be finite. However, one could try to argue as follows. It is clear from the Raychaudhuri equation that the introduction of NEC-violating matter will cause the right-hand side of the equation, and therefore the slope of $\theta$ as a function of $\lambda$, to become less negative. This will mean that the caustic will form at a larger value of $\lambda$, so the lightsheet may extend farther. In a situation in which the entropy bound is saturated or nearly saturated, the result could be a violation of the entropy bound. The problem with this kind of argument is that $\lambda$ is not related in any simple, general way to the distance to the caustic. In fact, the relationship between $\lambda$ and distance is influenced by the kind of matter we introduce, and so is the initial value of $\theta$. Hence, a larger value of $\lambda$ does not necessarily mean that the caustic has moved farther away, and so one cannot conclude directly that NEC-violating matter endangers the usual entropy bound [in terms of $A/4$]. Perhaps the usual bound is always protected in this way, though it seems more likely that the $A/4$ bound does have to be replaced by an expression depending both on $A$ and on some parameter which measures the amount of NEC-violating matter. For the present, the point we wish to stress is that, if a caustic still forms, the entropy is at least still {\em bounded}, so some kind of simple entropy bound might still exist.  

Alternatively, the last term in equation 4 may dominate and cause $\theta$ eventually to increase and become positive without first diverging to $-\infty$. This would still be perfectly acceptable, since the lightsheet would terminate when $\theta$ vanishes; {\em the entropy would still be bounded}, again perhaps by some expression involving the area and a parameter measuring NEC-violation. [Of course, if the density and pressure are not constant, we could have both kinds of behaviour, depending on the position of the starting point from which one uses the Raychaudhuri equation to evolve the expansion.] 

The problem lies in intermediate cases: for example, if the positive and negative terms in the right side of the Raychaudhuri equation both asymptotically approach constant values and cancel each other, then $\theta$ might asymptotically approach some fixed negative value or zero, without ever diverging or vanishing. That is, $\theta$ might remain permanently negative, and so the lightsheet might never terminate. [This could of course also happen in other ways if the stress-energy tensor is more complex --- for example, the final term in the Raychaudhuri equation might oscillate permanently, and so on.] Generically [that is, unless the geometry or the matter distribution is very special] that would mean that the entropy on the light sheet would not be finite, and so no entropy bound could be true. Obviously it would be a grave blow to the holographic programme if one could construct a physically well-motivated example in which a finite area corresponds to a volume containing an infinite amount of entropy.

It is clear from this discussion that it is not possible to make a general statement about what happens to the entropy bound when the NEC is violated. Perhaps the NEC-violating matter is such that $\theta$ either always diverges or always vanishes after a finite affine parameter value, in which case some kind of entropy bound might continue to make sense [if finite affine parameter corresponds to finite distance]. One might think that it would be difficult to construct an example of this kind, however. Surprisingly, guided by physical considerations, one can in fact do this quite easily. This is the subject of the next section.

\addtocounter{section}{3}
\section*{3.  An Example of a NEC-Violating Spacetime With Finite Entropy on Lightsheets}

In view of the above discussion, we propose that a good way to probe fundamental aspects of holography is to introduce NEC-violating matter into Anti-deSitter space, $AdS_5$. The hope is that whatever we learn in this way --- perhaps from the AdS/CFT correspondence --- will teach us something about the status of NEC-violating matter in cosmology.

Of course, there are many ways in which one might do this. As the only form of NEC-violating matter which is close to being observed is [possibly] the cosmological dark energy, it is natural to use a ``cosmological" model of the matter. In cosmology it is customary to model the matter content of the universe with matter having a {\em constant} [or piecewise constant] equation-of-state parameter $w$, simply because the data are not yet capable of detecting variations of $w$ with time. The simplicity of the resulting equation of state allows one \cite{kn:silbergleit1}\cite{kn:silbergleit2} to obtain an exact solution of the Einstein equation, and this again is a major advantage. Therefore, to the extent that the densities and pressures of the NEC-violating matter in our model can have a ratio differing from $-1$, we shall take the ratio to be a constant.

In order to obtain a concrete model, we must now choose a particular representation of $AdS_5$.
One of the most interesting representations of $AdS_5$ is obtained by foliating it by copies of $AdS_4$. If $g(AdS_n)$ is the metric on $AdS_n$, then this foliation is expressed by
\begin{equation} \label{eq:five}
g(AdS_5) = dr\otimes dr + cosh^2({{r}\over{L}})\,\,g(AdS_4),
\end{equation}
where $r$ is a coordinate which ranges from $-\infty$ to $+\infty$, and $L$ is the ``radius" of $AdS_n$. This representation of the metric was used by Karch and Randall \cite{kn:karch1} to investigate the ``locality of localization". By cutting off the $AdS_5$ bulk at some fixed value of $r$, one obtains a braneworld model with an $AdS_4$ brane. Of course, this is not even a quasi-realistic model [but see \cite{kn:karch2}] since the cosmological constant on the brane is negative; in order to make it more realistic, one might try to modify the matter content of the bulk in such a way that the negative cosmological constant on the brane relaxes to zero. Although the models they consider are quite different, the results of Cline and Firouzjahi \cite{kn:cline1}\cite{kn:cline2}suggest that this can probably only be done by using NEC-violating matter. Thus we have a very natural setting for investigating the holography of such matter: introduce it into the Karch-Randall model in such a way that the brane cosmological constant is reduced to zero. Of course our interest is in the holography of this situation, not in trying to produce a truly realistic variant of the Karch-Randall model.

We therefore begin with the familiar ansatz 
\begin{equation} \label{eq:six}
g(bulk) = dr\otimes dr + e^{2A(r)}\,\,\eta_{ij}dx^i\otimes dx^j,
\end{equation}
where $r$ is as above, where $A(r)$ is a function to be determined and $\eta_{ij}dx^i \otimes dx^j$ is the Minkowski metric. This is of course the standard metric ansatz for a bulk corresponding to a flat brane; see for example \cite{kn:warner}. Now if we consider forms of cosmological matter which are designed to have pressure $p$ and positive density $\rho$ such that $p < -\rho$ when all fields depend only on time, then we shall find that they behave in a similar way if we introduce them into a geometry with a metric of the form given by equation 6. For example, Caldwell \cite{kn:caldwell} considers a toy model of NEC-violating matter obtained simply by reversing the sign of the kinetic term in a standard scalar field Lagrangian with a positive potential. [See \cite{kn:kachru} for an example of this kind of behaviour arising in string theory; see also \cite{kn:melchiorri}\cite{kn:axel} for other approaches.] The [five-dimensional] stress tensor of such a field in the geometry given by equation 6 is 
\begin{equation} \label{eq:seven}
T_{\mu\nu} = -\partial_\mu\phi\partial_\nu\phi + {1\over 2}g_{\mu\nu}\partial_\alpha\phi\partial^\alpha\phi - g_{\mu\nu}V(\phi),
\end{equation}
and if we assume now that $\phi$ depends only on $r$ we find that the density is
\begin{equation} \label{eq:eight}
\rho = -{1\over2}(\phi')^2 + V(\phi),
\end{equation}
where $\phi'$ is the derivative of $\phi$ with respect to $r$. The pressure transverse to the brane is
\begin{equation} \label{eq:nine}
p_T = -{1\over2}(\phi')^2 - V(\phi),
\end{equation}
while the pressures parallel to the brane are all given by
\begin{equation} \label{eq:ten}
p_P = {1\over2}(\phi')^2 - V(\phi),
\end{equation}
which of course is just the negative of $\rho$, as required by Poincar\'e invariance in the direction parallel to the brane. We see at once that $\rho$ is always positive for sufficiently large potentials, but that $p_T \leq -\rho$. A similar result holds for the more complicated NEC-violating matter discussed by Melchiorri et al \cite{kn:melchiorri}. Thus, introducing such matter into a geometry given by equation 6, we will obtain in general a matter distribution with anisotropic pressures; the pressure is equal to minus the density $\rho$ in the direction parallel to the brane, and the transverse pressure $p_T$ is less than $-\rho$ .

As explained earlier, we shall follow cosmological practice and approximate the ``transversal equation of state" of the NEC-violating matter, whatever it may be, by
\begin{equation} \label{eq:eleven}
p_T = w\rho,
\end{equation}
where $w$ is a {\em constant} satisfying $w < -1$. This matter is to be superimposed on the $AdS_5$ background, with its density $-3/(4\pi L^2)$ and its pressure $+3/(4\pi L^2)$. From \cite{kn:warner} we have Einstein equations
\begin{equation} \label{eq:twelve}
-3A'' - 6(A')^2 = 8\pi ({{-3}\over {4\pi L^2}} + \rho),
\end{equation}
and
\begin{equation} \label{eq:thirteen}
6(A')^2 = 8\pi ({{3}\over {4\pi L^2}} + p_T),
\end{equation}
where $p_T$ is related to $\rho$ by equation 11. 

From equation 13 we have 
\begin{equation} \label{eq:fourteen}
8\pi {dp_T \over dr} = 12A'A'',
\end{equation}
and since 12 and 13 imply
\begin{equation} \label{eq:fifteen}
3A'' = -8\pi (\rho + p_T),
\end{equation}
we have
\begin{equation} \label{eq:sixteen}
{dp_T \over dr} = -4A'(\rho + p_T).
\end{equation}
Combining this with equation 11 we have
\begin{equation} \label{eq:seventeen}
{dp_T \over dA} = -4(1 + w^{-1})p_T.
\end{equation}
It is convenient now to define a parameter $\beta$ by
\begin{equation} \label{eq:eighteen}
\beta = 4(1 + w^{-1}).
\end{equation}
Clearly $\beta$ is a number satisfying $0 < \beta < 4$. Solving equation 17 we obtain
\begin{equation} \label{eq:nineteen}
p_T = -Pe^{-\beta A},
\end{equation}
where $P$ is a positive constant. Equation 13 now becomes
\begin{equation} \label{eq:twenty}
6(A')^2 = 8\pi ({{3}\over {4\pi L^2}} - Pe^{-\beta A}),
\end{equation}
and this is a differential equation to be solved for $A$.

This equation can be solved in terms of elementary functions: we have
\begin{equation} \label{eq:twentyone}
e^A = ({{4\pi L^2P} \over {3}})^{1/\beta}cosh^{\left({2/\beta}\right)}({{\beta r} \over {2L}}),
\end{equation}
and so our NEC-violating metric is
\begin{equation} \label{eq:twentytwo}
g(NECV) = dr \otimes dr + ({{4\pi L^2P} \over {3}})^{2/\beta}cosh^{\left({4/\beta}\right)}({{\beta r} \over {2L}})\,\,\eta_{ij}dx^i\otimes dx^j.
\end{equation}
By scaling the Minkowski coordinates on the brane, we can of course adjust the value of $P$ arbitrarily, to suit our convenience. One possible choice is $P = 3/4\pi L^2$, which is useful for studying the limit when $\beta$ tends to zero: since the limit of $cosh^{\left({4/\beta}\right)}({{\beta r} \over {2L}})$ is unity, we see that the metric becomes flat. This is reasonable, since in that limit we are simply using a positive cosmological constant to cancel the background $AdS_5$ negative cosmological constant, and of course the metric should be flat in that case. Another useful choice of $P$, the one we shall adopt henceforth, is obtained by noticing that for values of $r$ tending to $\infty$, the metric is approximately
\begin{equation} \label{eq:twentythree}
dr \otimes dr + ({{\pi L^2P} \over {3}})^{2/\beta}e^{(2r/L)}\,\,\eta_{ij}dx^i\otimes dx^j.
\end{equation}
Thus we regain precisely the standard $AdS_5$ metric near infinity if we choose $P = 3/\pi L^2$. Our metric is then
\begin{equation} \label{eq:twentyfour}
g(NECV) = dr \otimes dr + 4^{(2/\beta)}cosh^{\left({4/\beta}\right)}({{\beta r} \over {2L}})\,\,\eta_{ij}dx^i\otimes dx^j,
\end{equation}
and this discussion shows that our metric is asymptotic to the $AdS_5$ metric in a quite straightforward sense. Clearly the matter is concentrated near the ``throat" at $r = 0$; the density and pressure must approach the AdS values as we move towards the boundary. In fact we have
\begin{equation} \label{eq:twentyfive}
p_T + \rho = {{\Lambda\beta}\over {32\pi}}sech^2({{\beta r} \over {2L}}),
\end{equation}
where $\Lambda = -6/L^2$ is the cosmological constant of the background $AdS_5$. Thus indeed the NEC-violating combination $p_T + \rho$ attains its maximum magnitude $|{{\Lambda\beta}\over {32\pi}}|$ at $r=0$, and it increases towards zero as the boundary is neared.
 
Like the Karch-Randall metric, our metric is completely free of all singularities; and, as in that case, the form of the metric suggests that the boundary is disconnected; but this is not so. In both cases, the $r = constant$ slices are themselves infinitely large [$AdS_4$ for Karch and Randall, Minkowski space here] and they all intersect ``at infinity" in such a way that the boundary is in fact connected. Notice that, again as in the Karch-Randall case, the function $A$ satisfies $A'' > 0$; this is clear from equation 15. However, in the Karch-Randall case this was due to the negative cosmological constant on the brane --- that is, it was ultimately due to the fact that the Weak Energy Condition [WEC] is violated both on the brane and in the bulk. Here it is due simply to the fact that we are violating the NEC instead of the WEC. In fact, if the conjectural identification of the c-function given in \cite{kn:warner} is correct, then the AdS/CFT interpretation of NEC-violating matter in the bulk is precisely that the c-function of the CFT on the boundary should evolve in just the opposite way to the usual. Naturally, as is pointed out in \cite{kn:warner}, exotic matter in the bulk should correspond to exotic behaviour of the CFT on the boundary.

Now, following Karch and Randall, we set up a braneworld at a specific negative value of $r$, namely $r = -c$, where $c$ is positive. We cut away all smaller values of $r$ and reflect around the cut point. [Henceforth, however, we shall ignore the other half of the bulk.] Exactly as for Karch and Randall, we have a warp factor which decreases until $r=0$ is reached, where it ``bounces" and then increases towards infinity. The region of the bulk corresponding to any given finite region on the brane evidently has infinite volume. As was mentioned earlier, it is not entirely obvious that infinite volume implies infinite entropy, so we shall give a rough argument that there is indeed cause for concern here. Following Flanagan et al \cite{kn:flanagan}, we shall suppose that the entropy in a region of the bulk is roughly related to the amount of matter [measured by a suitable integral of $|T_{ab}k^ak^b|$] it contains. From equations 24 and 25  we see that the relevant integral is
\begin{equation} 
{{3A\beta}\over {16\pi L^2}}\int^\infty_{-c}\,\,{{cosh^{\left({{6}\over {\beta}}- 2\right)}({{\beta r} \over {2L}})dr} \over {cosh^{\left({{6}\over {\beta}}\right)}({{\beta c} \over {2L}})}},
\end{equation}
where $A$ is the three-dimensional ``area" of some region on the brane. This is actually finite for $\beta$ between 3 and 4, so we see that infinite volume does not necessarily imply an infinite amount of matter. However, realistic values of $\beta$ are no larger than about unity, so in the case of real interest to us it would indeed be disastrous if a lightsheet ever extended infinitely far into the bulk. We must show that this cannot happen.

The study of the lightsheets associated with the braneworld is very much simplified by the following simple observation. The metric given by equation 24 may be written as 
\begin{equation} 
g(NECV) = 4^{(2/\beta)}cosh^{\left({4/\beta}\right)}({{\beta r} \over {2L}})[ds \otimes ds + \eta_{ij}dx^i\otimes dx^j],
\end{equation}
where $s$ is a new coordinate defined by
\begin{equation} 
{{dr}\over {ds}} = 4^{(1/\beta)}cosh^{\left({2/\beta}\right)}({{\beta r} \over {2L}}).
\end{equation}
Evidently the metric is conformally flat. The equation governing the evolution of the shear away from the brane [see \cite{kn:wald}, page 223],
\begin{equation} 
k^a\nabla_a\sigma_{bc} = -\theta\sigma_{bc} + h^d_bh^e_cC_{jfeg}k^fk^g,
\end{equation}
where $h^a_b$ is the projection operator and $C_{abcd}$ is the Weyl tensor, shows that if the shear vanishes on the brane, then it will vanish everywhere in the bulk. So we can ignore the  shear term in the Raychaudhuri equation henceforth.

We now choose two null vectors, $k^a$ and $l^a$, perpendicular to the brane slice $r = -c$, $t = 0$. We take 
\begin{equation} 
k^t = 4^{(-1/\beta)}sech^{\left({2/\beta}\right)}({{\beta r} \over {2L}}),\:\: k^r = +1,\:\: k^x = k^y =   k^z = 0
\end{equation}
and
\begin{equation} 
l^t = 4^{(-1/\beta)}sech^{\left({2/\beta}\right)}({{\beta r} \over {2L}}),\:\: l^r = -1,\:\: l^x = l^y =  l^z = 0.
\end{equation}
Here $k^a$ corresponds to a ray of light moving into the bulk in the positive $r$ direction. Clearly $k^al_a = -2$, and the projection operator is 
\begin{equation} 
h^a_b = diag(\:0, \:0, \:1, \:1, \:1 \:).
\end{equation}
The null extrinsic curvature on the brane is then 
\begin{equation} 
K^a_b = {{-1}\over{L}}tanh({{\beta c} \over {2L}})h^a_b,
\end{equation}
and the expansion at the brane is
\begin{equation} 
\theta(-c) = {{-3}\over{L}}tanh({{\beta c} \over {2L}}).
\end{equation}
Now the flat metric
\begin{equation} 
ds \otimes ds + \eta_{ij}dx^i\otimes dx^j
\end{equation}
is obtained from $g(NECV)$, as given by equation 27, by means of the conformal transformation with conformal factor $\Omega$ given by
\begin{equation} 
\Omega^2 = 4^{(-2/\beta)}sech^{\left({4/\beta}\right)}({{\beta r} \over {2L}}).
\end{equation}
Therefore the canonical parameter $\tilde{\lambda}$ for null geodesics of the flat metric is obtained [see \cite{kn:wald}, page 446] from the canonical parameter for null geodesics of $g(NECV)$ by means of the simple formula
\begin{equation} \label{eq:thirtysix}
d\tilde{\lambda} = 4^{(-2/\beta)}sech^{\left({4/\beta}\right)}({{\beta r} \over {2L}})d\lambda.
\end{equation}
But since the metric 35 is flat, its null geodesics satisfy $ds/d\tilde{\lambda} = constant = C^*$. Therefore we have
\begin{equation} \label{eq:thirtyseven}
{{dr}\over {d\lambda}} = {{dr}\over {ds}}{{ds}\over {d\widetilde{\lambda}}}{{d\widetilde{\lambda}}\over {d\lambda }} =  (4^{(1/\beta)}cosh^{\left({2/\beta}\right)}({{\beta r} \over {2L}}))(C^*)(4^{(-2/\beta)}sech^{\left({4/\beta}\right)}({{\beta r} \over {2L}})) = Csech^{\left({2/\beta}\right)}({{\beta r} \over {2L}}),
\end{equation}
where $C$ is a constant.

Now ${{dr}\over {d\lambda}} = k^r = 1$ at the brane, so we have 
\begin{equation} \label{eq:thirtyeight}
C = cosh^{\left({2/\beta}\right)}({{\beta c} \over {2L}}).
\end{equation}
Hence
\begin{equation} \label{eq:thirtynine}
{{dr}\over {d\lambda}} = cosh^{\left({2/\beta}\right)}({{\beta c} \over {2L}})sech^{\left({2/\beta}\right)}({{\beta r} \over {2L}}).
\end{equation}

Returning now to the Raychaudhuri equation and dropping, as explained, the terms in the vorticity and shear, we have from equation 25
\begin{equation} \label{eq:forty}
{{d\theta}\over {d\lambda}} = {{-1}\over {3}}\theta^2 +
{{3\beta}\over {2L^2}}sech^2({{\beta r} \over {2L}}),
\end{equation}
and combining this with equation 40 we have
\begin{equation} \label{eq:fortyone}
cosh^{\left({2/\beta}\right)}({{\beta c} \over {2L}})sech^{\left({2/\beta}\right)}({{\beta r} \over {2L}}){{d\theta}\over {dr}} = {{-1}\over {3}}\theta^2 +
{{3\beta}\over {2L^2}}sech^2({{\beta r} \over {2L}}). 
\end{equation}
{\em This is the fundamental differential equation governing the evolution of the expansion as we move into the bulk.} The initial condition is given by equation 34. Bear in mind that, from equation 24, $r$ has a direct metric significance: it measures distances into the bulk from the brane.

Despite the simplicity of its form, equation 42 cannot be solved exactly; a numerical investigation is needed. Extensive trials reveal that the graphs of $\theta$ as a function of $r$ are always of two types, depending on the position of the brane [that is, the value of $c$.]
The initial value of the slope of $\theta(r)$ is given, from equations 34 and 42, by
\begin{equation} \label{eq:fortytwo}
{{d\theta}\over {dr}}(-c) = {{3\beta}\over {2L^2}}[1 - (1 +{{2}\over {\beta}})tanh^2({{\beta c} \over {2L}})].
\end{equation}
As one might expect, when the brane is relatively far from the throat [where, it will be remembered, the NEC-violating matter is most concentrated], then the expansion initially decreases; however, as a ray encounters more of the NEC-violating matter near to the throat, the expansion can begin to increase, or, if $c$ is large, remain almost constant for a while. Beyond the throat, however, it declines again, and the result is a caustic: the NEC-violating matter is unable to prevent this in this case. A graph of a typical example of $\theta(r)$ is given in Figure 1; notice that $c$ has been chosen to be substantially larger than $L$, which is what we mean by the brane being ``relatively far" from the throat. The value chosen for $\beta$, namely unity, corresponds by equation 18 to $w = -4/3$, a value compatible with the observational data \cite{kn:melchiorri} and also chosen as an illustrative value by Frampton \cite{kn:frampton} and by Dabrowski et al \cite{kn:dabrowski1}\cite{kn:dabrowski2}.

\begin{figure}[!h]
\centering
\includegraphics[width=0.8\textwidth]{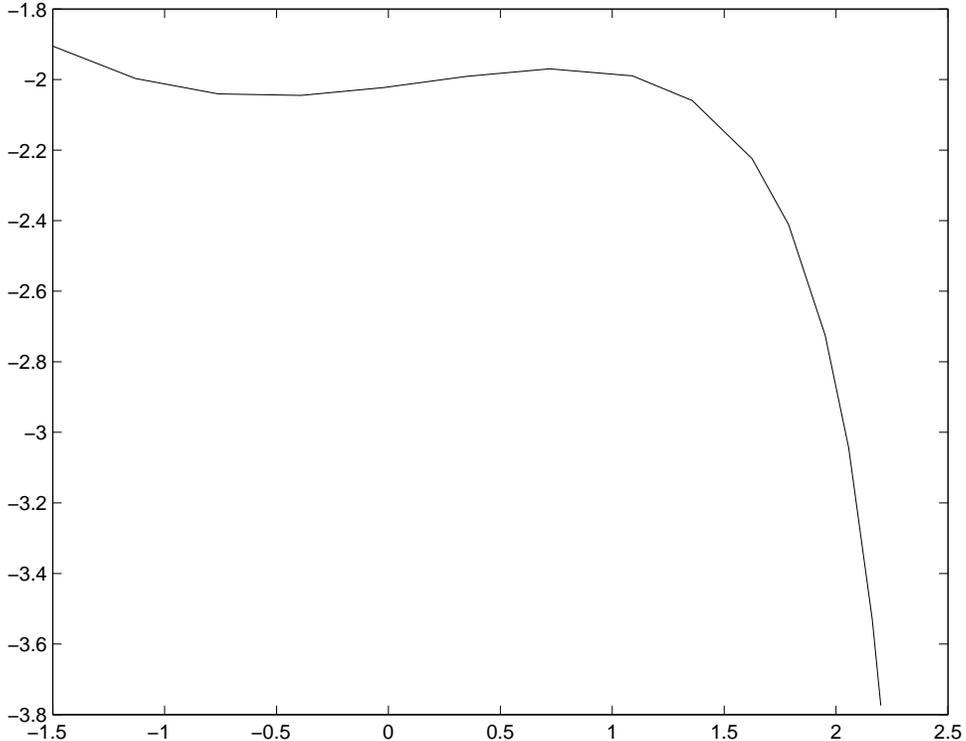}
\caption{$\theta(r)\:\:with\:\: \beta = L = 1,\: c = 1.5$}
\end{figure}

However, when the brane is relatively near to the throat, $\theta(r)$ immediately rises and reaches zero; it continues to increase for a while beyond the throat, before decreasing and tending asymptotically to zero from above. A typical example is shown in Figure 2, below: here $c < L$.

\begin{figure}[!h]
\centering
\includegraphics[width=0.8\textwidth]{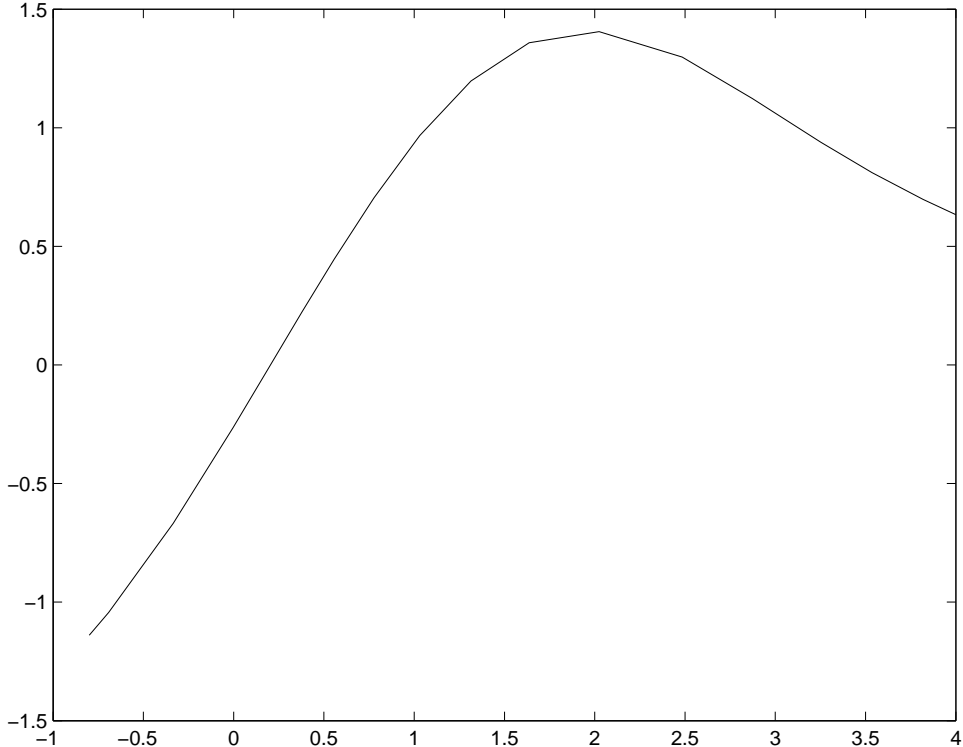}
\caption{$\theta(r)\:\: with\:\: \beta = L = 1,\: c = 0.8$}
\end{figure}

In both cases, the lightsheet terminates at a finite value of $r$, that is, at a finite distance from the braneworld. In no case does the lightsheet extend infinitely far into the bulk.

It would appear, then, that the NEC is [in this special case] not needed to ensure the finiteness of the entropy; the entropy on all lightsheets is finite, even though the NEC is violated everywhere in the bulk. The obvious next question is: is the entropy bound satisfied?  However, now that we have established that the ``holographic domain" [see below] of our brane is in fact always finite, another, even more basic question arises. We have to check that a holographic formulation of this situation is {\em internally consistent}.

\section*{4. The Bousso-Randall Consistency Condition}

If we consider the case illustrated in Figure 1, we see that there is a brane at $r = -1.5$, and a caustic on the other side of the throat, at about $r = 2.2$. Let us call this value $b$. The region corresponding to values of $r$ between the brane and the caustic is called \cite{kn:boussorandall} the {\em holographic domain} of the brane: it is the part of the bulk which is related to the brane by holography. [In situations like the one corresponding to Figure 2, $b$ is defined as the value of r where $\theta = 0$.] The region beyond the caustic [or the region where $\theta = 0$] is irrelevant to the brane. This led Bousso and Randall to construct a very simple yet stringent consistency check: they point out that this picture of holography can only really make sense if the brane holographic image of a bulk particle ``disappears" when the bulk particle leaves the holographic domain. They gave a very elegant demonstration that, in the case of the Karch-Randall brane, the holographic image moves out to the conformal infinity of the $AdS_4$ brane precisely when the particle reaches the distance of the caustic [slightly beyond the throat] from the brane. Thus the Karch-Randall model passes the consistency check.

The consistency check uses the concept of ``causal diamonds", which is explained very clearly in \cite{kn:boussorandall}; we refer the reader there for the details. Let $P$ be a point in the bulk of our NEC-violating spacetime, located at $ r = r(P)$, and let $q$ be a point on the brane which is reached by a ray of light from $P$. We think of $q$ as the tip of a five-dimensional causal diamond, the intersection of which with the brane should give a four-dimensional causal diamond. The event $P$ should correspond to a CFT excitation on the brane having support on the boundary of the four-dimensional causal diamond. Using equation 24, we have, along the null geodesic joining $P$ to $q$,
\begin{equation} \label{eq:fortythree}
(dr)^2 - 4^{(2/\beta)}cosh^{\left({4/\beta}\right)}({{\beta r} \over {2L}})\,\,(dt)^2 = 0.
\end{equation}
The holographic image of $P$ is a thin shell on the brane, of radius given by the distance light can cover in the time taken by the light to travel from $P$ to $q$. Since the geometry of the brane is Minkowskian, this radius is simply given by integrating $dt$ from equation 44 [while correcting for the fact that the Minkowski coordinates on the brane do not give proper times and distances according to the metric 24]:
\begin{equation} \label{eq:fortyfour}
R(P) = cosh^{\left({2/\beta}\right)}({{\beta c} \over {2L}})\int^{r(P)}_{-c}\,\, sech^{\left({2/\beta}\right)}({{\beta r} \over {2L}})\,\,dr.
\end{equation}
Bousso and Randall show that the corresponding formula for $R(P)$ in the Karch-Randall case diverges precisely when $r(P)$ reaches $b$, the value of $r$ at the edge of the holographic domain. This is the ``evanescence of the CFT shell" as the bulk particle leaves the holographic domain of the brane. But here, if $P$ moves to the edge, we have
\begin{equation} 
R(P_{edge}) = cosh^{\left({2/\beta}\right)}({{\beta c} \over {2L}})\int^{b}_{-c}\,\, sech^{\left({2/\beta}\right)}({{\beta r} \over {2L}})\,\,dr.
\end{equation}
and this is {\em always finite}. Thus, as the bulk particle moves far beyond the holographic domain of the brane, its ``CFT shell" is still very much present on the brane; it shows no sign of disappearing off to the conformal infinity of Minkowski space. To put it another way: one would expect a CFT shell that moves off to Minkowski conformal infinity to correspond to a particle moving beyond a horizon in the bulk, as happens for $AdS_5$ when it is foliated by Minkowski slices [in the familiar Poincar\'e coordinates]. But here we have no bulk horizon. In fact, no matter how far into the bulk the particle penetrates, the radius of the CFT shell can never exceed
\begin{equation} 
R_{max} = cosh^{\left({2/\beta}\right)}({{\beta c} \over {2L}})\int^{\infty}_{-c}\,\, sech^{\left({2/\beta}\right)}({{\beta r} \over {2L}})\,\,dr,
\end{equation}
which is finite; for the parameter values given in Figure 1, $R_{max}$ is approximately 5.48. A CFT shell of greater radius than this {\em simply has no bulk counterpart}, surely as direct a contradiction of holography as one could hope to find. 

Bousso and Randall also investigated the fate of a particle which moves beyond the holographic domain of the Karch-Randall brane. They found that shortly before the CFT shell leaves the braneworld, a second CFT shell appears on the surviving half of the $AdS_5$ boundary, so the particle already has this CFT shell at infinity when it loses its original CFT shell. [The particle thus briefly has {\em two} CFT shells, a curious fact with no obvious interpretation.] The holographic domain of the ``brane at infinity", that is, the surviving half of the $AdS_5$ boundary, is just the corresponding half of the $AdS_5$ bulk. In the present case, the situation is quite different: numerical experimentation strongly suggests that if we push the brane out towards $r = -\infty$, then its holographic domain protrudes farther into the region $r \geq 0$, so that $b$ tends to infinity. See for example Figure 3, below, where the brane is at $r = -30$ and $b$ is approximately 23.5. [It seems that, for large values of $c$, $b$ is always somewhat smaller than $c$ but of the same order of magnitude.]

\begin{figure}[!h]
\centering
\includegraphics[width=0.8\textwidth]{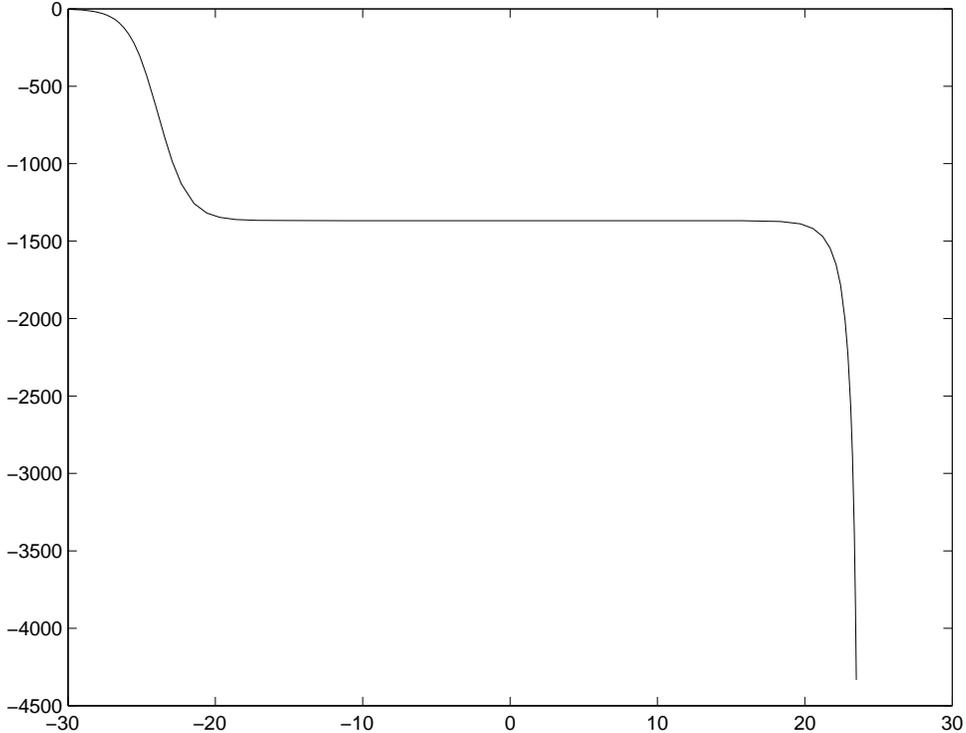}
\caption{$\theta(r)\:\: with\:\: \beta = L = 1,\: c = 30$}
\end{figure}

Thus, the holographic domain of a braneworld pushed out to $r = -\infty$ is the entire bulk. Again, however, there are CFT shells at $r = -\infty$ with no bulk counterpart. To see this, recall that there is no metric on this ``brane", only a conformal structure. Choose $\eta_{ij}dx^i\otimes dx^j$ as a representative of this conformal structure, and let $u$ be the corresponding radial coordinate $(x^2 + y^2 + z^2)^{1/2}$. [The reader should think of $u$ as nothing more than a {\em coordinate} which distinguishes points on the boundary but which has no meaning as a ``radius".] The value of this coordinate for the CFT shell of a particle at $r = r(P)$ is just 
\begin{equation} 
u(P) = \int^{r(P)}_{-\infty}\,\, sech^{\left({2/\beta}\right)}({{\beta r} \over {2L}})\,\,dr,
\end{equation}
and again this is strictly finite, even if $r(P)$ tends to infinity. Again, no CFT shell at $r = -\infty$ which corresponds to values of $u$ greater than $\int^{\infty}_{-\infty}\,\, sech^{\left({2/\beta}\right)}({{\beta r} \over {2L}})\,\,dr$ has any bulk counterpart. Of course all this applies equally to the ``brane" at $r = +\infty$, and so we obtain a picture of the ``global holography" \cite{kn:boussorandall} of this spacetime as follows. Any particle {\em anywhere} in the bulk is always associated with two CFT shells, one on the braneworld, and one at $r = +\infty$. The particle is always in the holographic domain of $r = +\infty$, but it can leave the holographic domain of the braneworld; however, nothing in particular happens to the CFT shells when this occurs. No CFT shell ever approaches the boundary of either the brane at $r = -c$ or the one at $r = +\infty$, and there are ``infinitely large" regions of both branes which have no bulk counterpart. Certainly no CFT shell ever crosses from one brane to the other, as happens in the Karch-Randall case: the two branes are disconnected in this sense.

In short, we can say that this spacetime is profoundly ``anti-holographic". A particle not in the holographic domain of the braneworld, which contributes nothing to the entropy on the brane lightsheets, is nevertheless detected by the CFT on the brane. On the other hand, there are infinitely many CFT shells, both on the braneworld and at $r = +\infty$, which correspond to nothing whatever in the bulk. If holography is correct, these are simply internal inconsistencies. Holography completely forbids a situation like this to arise.  

\section*{5.  Conclusion}

It is difficult to over-state the importance of the cosmological NEC. If indeed $w < -1$, then we may have to adapt ourselves to thinking of the Universe as an intrinsically unstable system, 
just as an earlier generation had to learn to think of the Universe as being dynamic rather than static. Furthermore, it is not easy to reconcile a violation of the NEC with basic physical principles: if $w < -1$, then matter on the largest scales is governed by radically new physics \cite{kn:caldwell}. The fate of the entire Universe could be controlled by quantum-gravitational effects acting not in the deep past, but in the here and now \cite{kn:starobinsky}. The Universe may be fated to end, not in a Crunch but in a Smash \cite{kn:caldwell}\cite{kn:mcinnessmash} [that is, a final singularity in which the cosmological scale factor tends to infinity in a finite proper time.]

Alternatively, it may be \cite{kn:guzzo} that the Universe does not violate the NEC. If so, the fact remains \cite{kn:melchiorri} that it comes perilously close to doing so, and it is essential to understand what prevents values of $w$ close to $-1$ from becoming values slightly less than $-1$. It may seem a great extrapolation to conclude from the example studied here that it is holography that plays this role. However, we wish to draw attention not just to the fact that holography forbids violations of the NEC in this particular example, but also to the {\em way} in which the prohibition is expressed. Holography, in this example, does not just predict something strange when NEC-violating matter is introduced into $AdS_5$: in fact, it fails to predict anything at all; it becomes completely incoherent; it claims both that the brane is aware of an excitation beyond the edge of the lightsheet, and that it is not. It may help to compare this with classical General Relativity. If we introduce unusual forms of matter into Minkowski space, then the theory may predict very strange spacetime structure. But if we tried to claim that we had introduced into Minkowski space a form of matter which is not locally conserved, then General Relativity tells us nothing, because of the Bianchi identities. The situation that results when NEC-violating matter is introduced into $AdS_5$ is similar. 

This suggests to us that holography forbids all violations of the cosmological version of the NEC. [Note that it was pointed out in \cite{kn:brustein} that the entropy bound can be used to {\em replace} energy conditions in cosmology. The present claim may be regarded as an attempt to explain that observation.] To prove this, one would have to consider NEC-violating matter with a general equation of state: that is, one would have to allow $w$ to depend on position. Further work is needed on this problem, but it seems implausible that this will materially change our conclusions. It is hard to believe that tinkering with $w$ will cause $R_{max}$ [given by equation 47] to become infinite, as it must if all CFT shells on the braneworld are to have holographic partners. Intuitively, the conflict with holography arises from the tendency of NEC-violating spacetimes to ``flare out" in the manner of a wormhole \cite{kn:visserworm}, and we therefore expect the pathologies we have seen in this example to persist for any NEC-violating bulk. If this is so, then astrophysical analyses of the value of $w$ are direct observational tests of holography, which apparently {\em predicts} that $w \geq -1$. 

\section*{Acknowledgements}
The author is grateful to Dr Lin Ping for help with the numerical analysis of equation 42, and to Soon Wanmei for helpful discussions.


\begin{thebibliography}{18}
\linespread{0.5}


\bibitem{kn:bousso1}
Raphael Bousso, A Covariant Entropy Conjecture, JHEP 9907 (1999) 004, arXiv:hep-th/9905177 
\bibitem{kn:bousso}
Raphael Bousso, The Holographic Principle, Rev.Mod.Phys. 74 (2002) 825, arXiv:hep-th/0203101 
\bibitem{kn:visserworm}
M. Visser, Lorentzian Wormholes, American Institute of Physics, 1995
\bibitem{kn:carroll}
Sean M. Carroll, The Cosmological Constant, Living Rev.Rel. 4 (2001) 1, arXiv:astro-ph/0004075    
\bibitem{kn:vissertwilight}
Carlos Barcelo, Matt Visser, Twilight for the energy conditions?, arXiv:gr-qc/0205066 
\bibitem{kn:mcinnessmash}
Brett McInnes, The dS/CFT Correspondence and the Big Smash, JHEP 0208 (2002) 029, arXiv:hep-th/0112066 
\bibitem{kn:mcinnesastro}
Brett McInnes, What If $w < -1$ ?, arXiv:astro-ph/0210321
\bibitem{kn:axel2}
Axel de la Macorra, Hector Vuccetich, Causality, Stability and Sound Speed in Scalar Field Models, arXiv:astro-ph/0212302
\bibitem{kn:caldwell}
R.R. Caldwell, A Phantom Menace? Cosmological consequences of a dark energy component with super-negative equation of state, Phys.Lett. B545 (2002) 23, arXiv:astro-ph/9908168 
\bibitem{kn:ratra}
P. J. E. Peebles, Bharat Ratra, The Cosmological Constant and Dark Energy, arXiv:astro-ph/0207347 
\bibitem{kn:frampton}
P.H. Frampton, How to Test Stringy Dark Energy, arXiv:astro-ph/0209037 
\bibitem{kn:mersini}
Mar Bastero-Gil, Paul H. Frampton, Laura Mersini, Modified Dispersion Relations from Closed Strings in Toroidal Cosmology, Phys.Rev. D65 (2002) 106002, arXiv:hep-th/0110167
\bibitem{kn:vafa}
R. Brandenberger, C. Vafa, Superstrings in the early universe, Nuclear Phys. B 316 (1989) 391
\bibitem{kn:uzan}
Martin Lemoine, Jerome Martin, Jean-Philippe Uzan, Trans-Planckian Dark Energy?, arXiv:hep-th/0212027 
\bibitem{kn:ellis}
S.W.Hawking, G.F.R. Ellis, The large scale structure of space-time, Cambridge University Press 1973  
\bibitem{kn:guth}
Arvind Borde, Alan H. Guth, Alexander Vilenkin, Inflation is not past-eternal, arXiv:gr-qc/0110012 
\bibitem{kn:boisseau}
B. Boisseau, G. Esposito-Farese, D. Polarski, A.A. Starobinsky, Reconstruction of a scalar-tensor theory of gravity in an accelerating universe, Phys.Rev.Lett. 85 (2000) 2236, arXiv:gr-qc/0001066.
\bibitem{kn:visser}
Carlos Barcelo, Matt Visser, Scalar fields, energy conditions, and traversable wormholes, Class.Quant.Grav. 17 (2000) 3843, arXiv:gr-qc/0003025 
\bibitem{kn:faraoni}
V. Faraoni, Superquintessence, Int.J.Mod.Phys. D11 (2002) 471, arXiv:astro-ph/0110067.
\bibitem{kn:torres}
Diego F. Torres, Quintessence, super-quintessence and observable quantities in Brans-Dicke and non-minimally coupled theories, Phys.Rev. D66 (2002) 043522, arXiv:astro-ph/0204504 
\bibitem{kn:arias}
Olga Arias, Tame Gonzalez, Israel Quiros, New exact cosmological solutions to Einstein's gravity minimally coupled to a Quintessence field, arXiv:gr-qc/0210097 
\bibitem{kn:olum}
Noah Graham, Ken D. Olum, Negative Energy Densities in Quantum Field Theory With a Background Potential, arXiv:hep-th/0211244 
\bibitem{kn:ford}
L.H. Ford, Adam D. Helfer, Thomas A. Roman, Spatially Averaged Quantum Inequalities Do Not Exist in Four-Dimensional Spacetime, arXiv:gr-qc/0208045  
\bibitem{kn:fewster}
C.J. Fewster, T.A. Roman, Null energy conditions in quantum field theory, arXiv:gr-qc/0209036 
\bibitem{kn:parker}
Leonard Parker, William Komp, Daniel A.T. Vanzella, Cosmological Acceleration Through Transition to Constant Scalar Curvature, arXiv:astro-ph/0206488 
\bibitem{kn:onemli}
V.K. Onemli, R.P. Woodard, Super-Acceleration from Massless, Minimally Coupled $\phi^4$, Class.Quant.Grav. 19 (2002) 4607, arXiv:gr-qc/0204065 
\bibitem{kn:bojowald}
Martin Bojowald, Inflation from Quantum Geometry, arXiv:gr-qc/0206054 
\bibitem{kn:turok}
Paul J. Steinhardt, Neil Turok, The Cyclic Universe: An Informal Introduction, arXiv:astro-ph/0204479 
\bibitem{kn:seiberg}
Nathan Seiberg, From Big Crunch To Big Bang - Is It Possible?, arXiv:hep-th/0201039 
\bibitem{kn:noh}
J. Hwang, H. Noh, Non-singular big-bounces and evolution of linear fluctuations, Phys.Rev. D65 (2002) 124010, arXiv:astro-ph/0112079 
\bibitem{kn:pintoneto}
Patrick Peter, Nelson Pinto-Neto, Primordial perturbations in a non singular bouncing universe model, Phys.Rev. D66 (2002) 063509, arXiv:hep-th/0203013 
\bibitem{kn:brandenberger}
Shinji Tsujikawa, Robert Brandenberger, Fabio Finelli, On the Construction of Nonsingular Pre-Big-Bang and Ekpyrotic Cosmologies and the Resulting Density Perturbations, Phys.Rev. D66 (2002) 083513, arXiv:hep-th/0207228 
\bibitem{kn:myung}
Y.S. Myung, Bouncing and cyclic universes in the charged AdS bulk background, arXiv:hep-th/0208086 
\bibitem{kn:sahni}
Ujjaini Alam, Varun Sahni, Supernova Constraints on Braneworld Dark Energy, arXiv:astro-ph/0209443 
\bibitem{kn:dabrowski1}
Mariusz P. Dabrowski, Wlodzimierz Godlowski, Marek Szydlowski, Brane universes tested by supernovae, arXiv:astro-ph/0210156
\bibitem{kn:dabrowski2}
Mariusz P. Dabrowski, Wlodzimierz Godlowski, Marek Szydlowski, Astronomical tests for brane universes and the dark energy on the brane, arXiv:astro-ph/0212100 
\bibitem{kn:kachru}
S. Kachru, L. McAllister, Bouncing Brane Cosmologies from Warped String Compactifications, arXiv:hep-th/0205209 
\bibitem{kn:hannestad}
Steen Hannestad, Edvard M\"ortsell, Probing the dark side: Constraints on the dark energy equation of state from CMB, large scale structure and Type Ia, Phys.Rev. D66 (2002) 063508, arXiv:astro-ph/0205096 
\bibitem{kn:lewis}
Geraint F. Lewis, Rodrigo A. Ibata, An investigation of gravitational lens determinations of $H_o$ in quintessence cosmologies, arXiv:astro-ph/0206425 
\bibitem{kn:dicus}
Duane A. Dicus, Wayne W. Repko, Discriminating between models for the dark energy, arXiv:hep-ph/0211109 
\bibitem{kn:magueijo}
Joao Magueijo, Levon Pogosian, Could thermal fluctuations seed cosmic structure?, arXiv:astro-ph/0211337 
\bibitem{kn:krauss}
Lawrence M. Krauss, Dark Energy and the Hubble Age, arXiv:astro-ph/0212369 
\bibitem{kn:guzzo}
Peter Schuecker, Robert R. Caldwell, Hans Böhringer, Chris A. Collins, Luigi Guzzo, Observational Constraints on General Relativistic Energy Conditions, Cosmic Matter Density and Dark Energy from X-Ray Clusters of Galaxies and Type-Ia Supernovae, arXiv:astro-ph/0211480 
\bibitem{kn:melchiorri}
Alessandro Melchiorri, Laura Mersini, Carolina J. Odman, Mark Trodden, The State of the Dark Energy Equation of State, arXiv:astro-ph/0211522 
\bibitem{kn:carrollmersini}
Sean M. Carroll, Laura Mersini, Can we live in a self-tuning universe?, Phys.Rev. D64 (2001) 124008, arXiv:hep-th/0105007
\bibitem{kn:cline1}
James M. Cline, Hassan Firouzjahi, No-Go Theorem for Horizon-Shielded Self-Tuning Singularities,
 Phys.Rev. D65 (2002) 043501, arXiv:hep-th/0107198
\bibitem{kn:cline2}
James M. Cline, Hassan Firouzjahi, Quest for a Self-Tuning Brane-World Solution to the Cosmological Constant Problem, arXiv:hep-th/0207155
\bibitem{kn:odintsov}
Emilio Elizalde, Shin'ichi Nojiri, Sergei D. Odintsov, Sachiko Ogushi, Casimir effect in de Sitter and Anti-de Sitter braneworlds, arXiv:hep-th/0209242 
\bibitem{kn:karch1}
Andreas Karch, Lisa Randall, Locally Localized Gravity, JHEP 0105 (2001) 008, arXiv:hep-th/0011156 
\bibitem{kn:porrati}
Massimo Porrati, Mass and Gauge Invariance IV (Holography for the Karch-Randall Model), Phys.Rev. D65 (2002) 044015, arXiv:hep-th/0109017 
\bibitem{kn:boussorandall}
Raphael Bousso, Lisa Randall, Holographic Domains of Anti-de Sitter Space, JHEP 0204 (2002) 057, arXiv:hep-th/0112080 
\bibitem{kn:wald}
R.M.Wald, General Relativity, University of Chicago Press, 1984
\bibitem{kn:silbergleit1}
A.S. Silbergleit, On cosmological evolution with the Lambda-term and any linear equation of state, arXiv:astro-ph/0208481 
\bibitem{kn:silbergleit2}
A.S. Silbergleit, On the Equation of State for Scalar Field, arXiv:astro-ph/0208481 
\bibitem{kn:karch2}
Andreas Karch, Lisa Randall, Localized Gravity in String Theory, Phys.Rev.Lett. 87 (2001) 061601, arXiv:hep-th/0105108  
\bibitem{kn:warner}
D.Z. Freedman, S.S. Gubser, K. Pilch, N.P. Warner, Renormalization Group Flows from Holography--Supersymmetry and a c-Theorem, Adv.Theor.Math.Phys. 3 (1999) 363, arXiv:hep-th/9904017
\bibitem{kn:axel}
A. de la Macorra, H. Vuccetich, Cosmology for Scalar Fields with $w<-1$ and Negative Potentials, arXiv:astro-ph/0212148 
\bibitem{kn:flanagan}
Eanna E. Flanagan, Donald Marolf, Robert M. Wald, Proof of Classical Versions of the Bousso Entropy Bound and of the Generalized Second Law, Phys.Rev. D62 (2000) 084035, arXiv:hep-th/9908070 
\bibitem{kn:starobinsky}
A.A. Starobinsky, Future and Origin of our Universe: Modern View, Grav.Cosmol. 6 (2000) 157, arXiv:astro-ph/9912054
\bibitem{kn:brustein}
Ram Brustein, Stefano Foffa, Avraham E. Mayo, Causal Entropy Bound for Non-Singular Cosmologies,
Phys.Rev. D65 (2002) 024004, arXiv:hep-th/0108098 

\end{thebibliography}
\end{document}